\documentclass[a4paper,fleqn]{cas-sc}

\usepackage[square,numbers]{natbib}
\usepackage{balance}
\usepackage{color,soul}
\usepackage{graphicx}
\usepackage{amstext}
\usepackage{amsmath}
\usepackage{amssymb}
\usepackage{booktabs}
\usepackage{caption}
\usepackage{subcaption}
\usepackage{hyperref}
\usepackage{url}
\usepackage{multirow}
\usepackage{multicol}
\usepackage{algorithm2e}
\usepackage{academicons}
\usepackage{xcolor}
\usepackage{xspace}
\def\tsc#1{\csdef{#1}{\textsc{\lowercase{#1}}\xspace}}
\tsc{WGM}
\tsc{QE}
\tsc{EP}
\tsc{PMS}
\tsc{BEC}
\tsc{DE}

\begin{document}
\let\WriteBookmarks\relax
\def\floatpagepagefraction{1}
\def\textpagefraction{.001}

\shorttitle{Med-IC}

\title [mode = title]{Med-IC: Fusing a Single Layer Involution with Convolutions for Enhanced Medical Image Classification and Segmentation} 


\tnotemark[]



\shortauthors{Islam et al.}

                   
\author[1]{Md. Farhadul Islam}[orcid=0000-0003-3249-4490]
\cormark[1]
\ead{farhadul.islam@bracu.ac.bd}
\author[1]{Sarah Zabeen}[orcid=0000-0002-9406-3816]
\ead{sarah.zabeen@bracu.ac.bd}
\author[1, 5]{Meem Arafat Manab}[orcid=0000-0002-2336-4160]
\ead{meem.arafat@bracu.ac.bd}
\author[2]{Mohammad Rakibul Hasan Mahin}[]
\ead{mohammad.rakibul.hasan.mahin@g.bracu.ac.bd}
\author[1, 4]{Joyanta Jyoti Mondal}[orcid=0000-0003-3113-8603]
\ead{joyanta@udel.edu}
\author[2]{Md. Tanzim Reza}[orcid=0000-0001-8964-1565]
\ead{tanzim.reza@bracu.ac.bd}
\author[6]{Md Zahidul Hasan}[]
\ead{h_mdzah@live.concordia.ca}
\author[3]{Munima Haque}[orcid=0000-0002-6712-6447]
\ead{munima.haque@bracu.ac.bd}
\author[2]{Farig Sadeque}[orcid=0000-0001-6797-7826]
\ead{farig.sadeque@bracu.ac.bd}
\author[1]{Jannatun Noor}[orcid=0000-0001-9669-151X]
\ead{jannatun.noor@bracu.ac.bd}

\cortext[cor1]{Corresponding Author}

\address[1]{Computing for Sustainability and Social Good (C2SG) Research Group, Department of Computer Science and Engineering, School of Data and Sciences, BRAC University, Dhaka, Bangladesh}
\address[2]{Department of Computer Science and Engineering, School of Data and Sciences, BRAC University, Dhaka, Bangladesh}
\address[3]{Biotechnology Program, Department of Mathematics and Natural Sciences, School of Data and Sciences, BRAC University, Dhaka, Bangladesh}
\address[4]{Department of Computer and Information Sciences, University of Delaware, United States of America}
\address[5]{School of Law and Government, Dublin City University, Ireland}
\address[6]{Concordia Institute for Information Systems Engineering (CIISE), Gina Cody School of Engineering and Computer Science, Concordia University, Canada}

\begin{abstract}
The majority of medical images, especially those that resemble cells, have similar characteristics. These images, which occur in a variety of shapes, often show abnormalities in the organ or cell region. The convolution operation possesses a restricted capability to extract visual patterns across several spatial regions of an image. The involution process, which is the inverse operation of convolution, complements this inherent lack of spatial information extraction present in convolutions. In this study, we investigate how applying a single layer of involution prior to a convolutional neural network (CNN) architecture can significantly improve classification and segmentation performance, with a comparatively negligible amount of weight parameters. The study additionally shows how excessive use of involution layers might result in inaccurate predictions in a particular type of medical image. According to our findings from experiments, the strategy of adding only a single involution layer before a CNN-based model outperforms most of the previous works. 

\end{abstract}



\begin{keywords}
Medical Imaging \sep
Involutional Neural Networks\sep
Convolutional Neural Networks \sep
Image Classification \sep
Medical Image Segmentation

\end{keywords}

\maketitle
\section{Introduction}
\label{sec:intro}

Medical image classification and segmentation are two fundamental tasks in computer-aided diagnostics. The accurate detection of diseases or abnormalities from images holds significant importance, facilitating phases including early diagnosis, treatment plans, and disease progression monitoring. CNN-based models have been dominant in these tasks \cite{JIANG2023106726}, although convolution-heavy models can be very costly with their large number of weight parameters \cite{alzubaidi_2021_review}. Medical image analysis, at times, requires effective spatial feature extraction to detect damaged tissues or anomalies that exhibit heterogeneous spatial patterns, often varying based on their location within an image \cite{pr1, pr2, pr3}. Therefore, models should ideally be able to identify features also based on their position, emphasizing the importance of spatial awareness. This understanding of spatial relationships has been shown to enhance diagnostic performance and contribute to increased realism in computer-aided diagnosis \cite{HOSSEINI2024100357, JAVED2021102104}.




\begin{figure*}[h]
    \centering

        \includegraphics[width=1\linewidth]{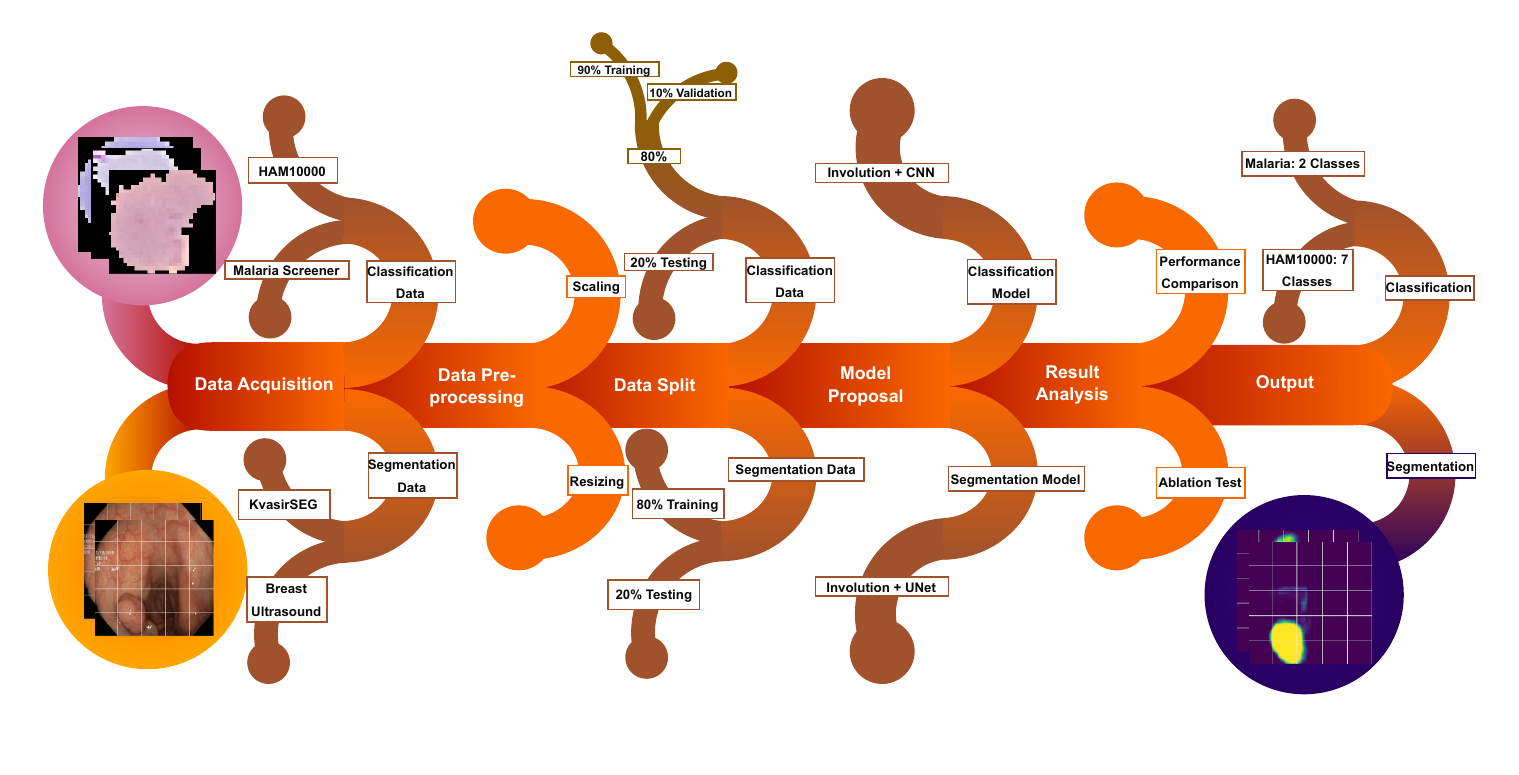}%
   
        \caption{Workflow of the proposed approach}
    \label{workflow}
        
\end{figure*}

Involution process \cite{involution_paper}, an offshoot of convolution, can be useful in this case. It utilizes fewer weight parameters and, with a dynamic kernel, extracts locational information, which is missing the convolution process. As a linear model, involution, much like attention layers, detects inter-positional dependencies quite well \cite{involution_paper}. Requiring neither a multitude of kernels nor position-wise variants, a single set of meta-weights is utilized on involution to reconstruct position-wise kernels. Compared to convolution, involution comes with a smaller number of weight parameters. This aids in constructing larger models. Additionally, the addition of involution layers to convolution models increases the number of weight parameters very slightly. The model does not necessarily become larger in terms of memory while we observe an increase in performance. 

The use of involution in segmentation has still not been explored in the medical imaging domain. In our study, we observe that using one involution layer can drastically impact performance. The dynamic kernel for each pixel is instantiated based on the value of the pixel and learning parameters. Involution overcomes the challenge of simulating long-range interactions by dynamically creating filters at each spatial position and functioning in a larger spatial arrangement, depending on the neighborhood. But, we also observe that overuse of this process typically leads to overfitting issues, since the images may have other spatial features that are not important for the task and involution may instead lead to an over-accrual of features. Therefore, minimal use of involution layers is suggested, and convolution still plays a crucial role. The diagnostic information included in histopathology images is critical, necessitating the extraction of global information first, followed by detailed extraction using conventional convolution. Our work demonstrates the use and the limitations of embedding involution layers in CNN-based model pipelines for medical image recognition tasks. 

Overall, we show how one involution layer can enhance both classification and segmentation tasks while also reducing the architecture's size. Moreover, this also leads to increased accuracy and recall. In this study, our key contributions include:

\begin{itemize}
    \item Proposing a novel architectural concept for classification and segmentation tasks in medical imaging.
    \item Discussing and analyzing how our proposed approach of adding one involution layer works and performs, where we illustrate involution kernel maps and GradCAM visualizations.
    \item Our proposed concept achieves excellent performance and outperforms previous methodologies. 
\end{itemize}

\begin{figure*}[h]
    \makebox[\linewidth]{
        \includegraphics[width=1\linewidth]{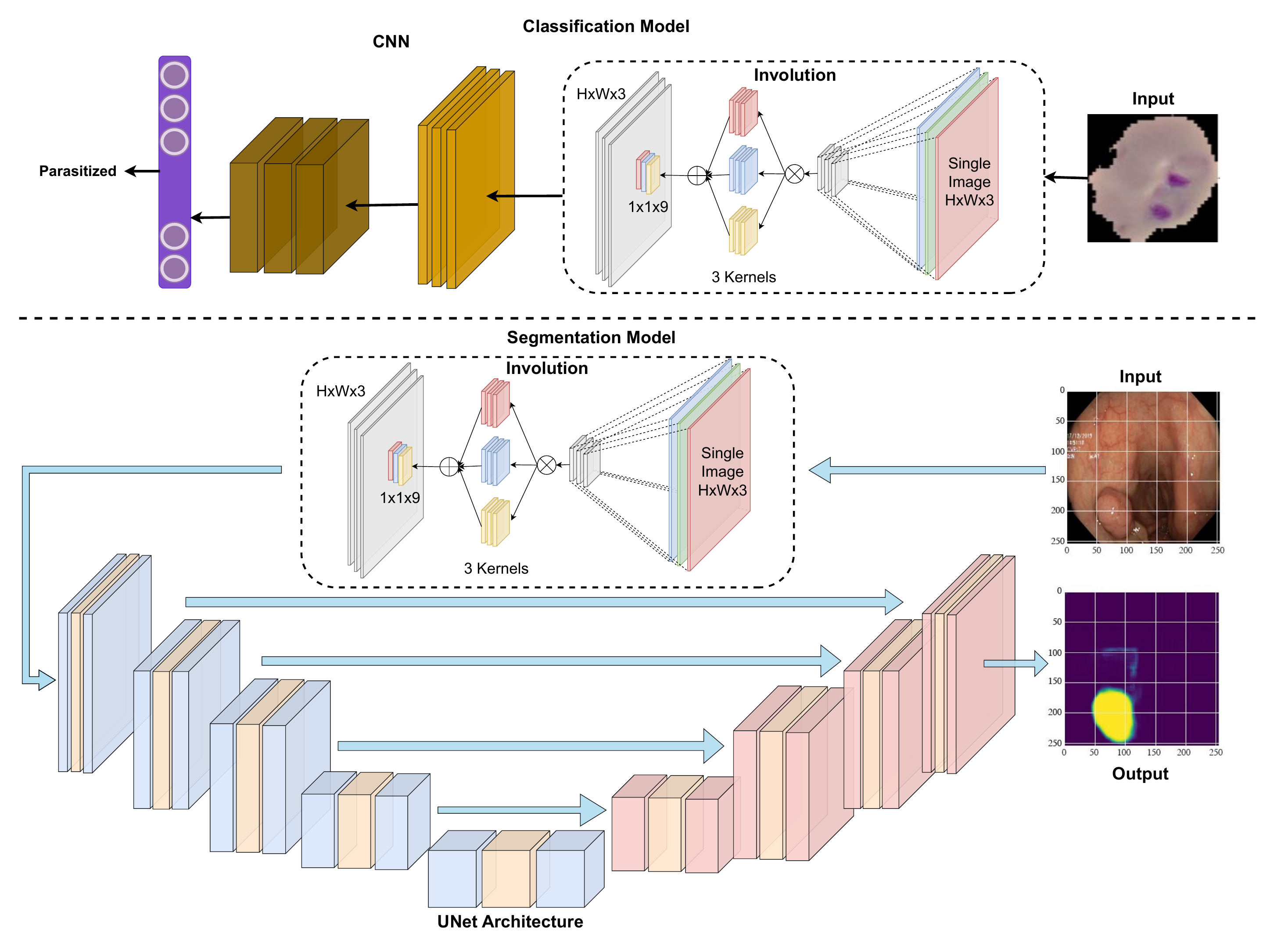}}

    \caption{Simplified illustration of our proposed architectural design. We add a single involution block before CNN-based models for important spatial feature extraction}
    \label{model}
\end{figure*}

\section{Literature Review}
\label{sec:li rev}

Convolutional Neural Networks (CNN) have evolved into a powerful image classifier in the past few years, which is why many researchers are now opting to work with it. CNN has proved to be useful in classifying skin cancer lesion types and detecting malaria-parasitized cells \cite{derm6, id12, id39, id40, id43}. Most of the works utilize different variations of convolution-based models such as MobileNet, ResNet, DenseNet, etc. For skin lesion classification HAM10000 \cite{ham} and ISIC \cite{CASSIDY2022102305} datasets have been used the most as they are the benchmark datasets in this task.

In medical image segmentation convolution-based U-Net variants have prevailed to work the best \cite{9446143, 10643318}. Other than these architectures, many other models have shown their capabilities. Ensembling with PraNet, CaraNet, and FCBFormer \cite{BOBOWICZ2024105522} for a computer-aided diagnosis system for breast cancer, dense residual U-Net for sclerosis lesions segmentation \cite{SARICA2023104965}. Other methodologies without the dominance of convolutions have also been studied and used in different medical image segmentation tasks \cite{NIDA201937, GUO2019105, Rahman_2023_WACV}.

Involution has been successful in various tasks, especially in medical imaging and vision tasks where spatial context is important \cite{10115109, islam2024involution, ICNet, 10288488, TIAN2022487}. Particularly concerning medical image classification, Islam et al. \cite{10115109} show the power of area-specific processes which are expertly combined with the reliable structure of convolutions, producing incredible effectiveness for the network. This is mainly evident in processes regarding cell-like images. However, it also mentions that the multiple involution layers lead to overfitting issues. They mainly focus on the reliability of this idea in the classification task. In our study, we extend this experiment and further prove why this is also true for both classification and segmentation in terms of resource efficiency.

The combination of involution and convolution is quite effective in medical image classification tasks but the idea of combining these two is quite new and according to our literature study, there are a limited number of works \cite{9948278, 9665787, 10112364}. Firstly, Gao et al. \cite{ICObj} presented a convolution-involution hybrid poised to detect monocular 3D objects. Similarly, Liang et al. \cite{ICNet} introduced the I-C Net, another instance of an involution-convolution hybrid, which functions more as a general classification model. However, both models differ in either design or implementation compared to ours. Our approach targets very particular image characteristics and incorporates uncertainty analysis and measurement, which is crucial for ensuring dependable predictions.

\section{Methodology}

The workflow in Figure \ref{workflow} outlines the step-wise tasks of our study. We first acquire benchmark datasets and pre-process them for optimal training. Next, we split the data, propose the model, run experiments, and obtain the outputs. Finally, we compare and analyze the results. Each step shows two branches for classification and segmentation tasks.

\subsection{Merging Involution with Convolution}

\begin{table}[]
\centering
\caption{Classification performance comparison of different variations of the hybrid models}
\begin{tabular}{c|c|ccc|ccc}
\hline
\multirow{2}{*}{Model} & \multirow{2}{*}{Weight Parameters $\downarrow$} & \multicolumn{3}{c|}{HAM10000}                                          & \multicolumn{3}{c}{Malaria}                                            \\ \cline{3-8} 
                       &                                       & \multicolumn{1}{c|}{Accuracy $\uparrow$} & \multicolumn{1}{c|}{Recall $\uparrow$} & F1 $\uparrow$ & \multicolumn{1}{c|}{Accuracy $\uparrow$} & \multicolumn{1}{c|}{Recall $\uparrow$} & F1 $\uparrow$ \\ \hline
Hybrid-1    or Med-IC (Cls)                & 301,983                             & \multicolumn{1}{c|}{98.85}    & \multicolumn{1}{c|}{99.41}  &    99.11      & \multicolumn{1}{c|}{98.67}    & \multicolumn{1}{c|}{98.55}  &    98.45      \\
Hybrid-2                   & 302,026                             & \multicolumn{1}{c|}{98.44}    & \multicolumn{1}{c|}{99.12}  &    98.44      & \multicolumn{1}{c|}{95.34}    & \multicolumn{1}{c|}{93.33}  &    93.33      \\
Hybrid-3                 & 302,069                             & \multicolumn{1}{c|}{96.67}    & \multicolumn{1}{c|}{95.94}  &    96.14      & \multicolumn{1}{c|}{92.76}    & \multicolumn{1}{c|}{91.29}  &     91.54    \\
CNN (3 Layers)         & 532,584                             & \multicolumn{1}{c|}{97.14}    & \multicolumn{1}{c|}{96.67}  &    96.67      & \multicolumn{1}{c|}{95.14}    & \multicolumn{1}{c|}{93.98}  &    94.89      \\
INN (1 Layer)   & 245,279                               & \multicolumn{1}{c|}{93.2}     & \multicolumn{1}{c|}{92.28}  &    92.50      & \multicolumn{1}{c|}{93.89}    & \multicolumn{1}{c|}{93.12}  &   93.69       \\ \hline
\end{tabular}
\label{abla_cls}
\end{table}

\begin{figure*}[h!]
    \makebox[\linewidth]{
        \includegraphics[width=0.65\linewidth]{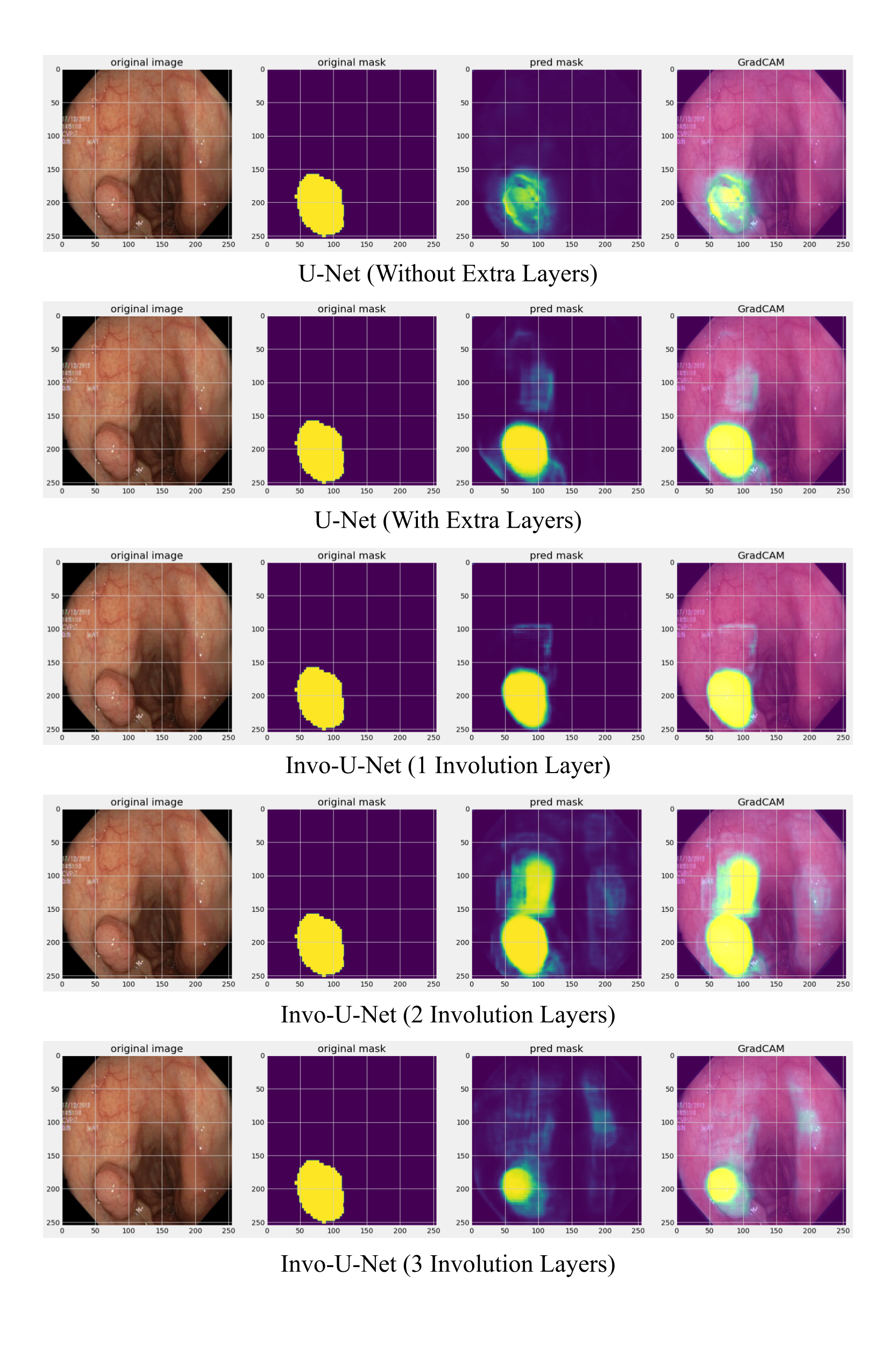}}

    \caption{Segmentation visualizations of Kvasir dataset}
    \label{res}
\end{figure*}

\begin{figure*}[h!]
    \makebox[\linewidth]{
        \includegraphics[width=.65\linewidth]{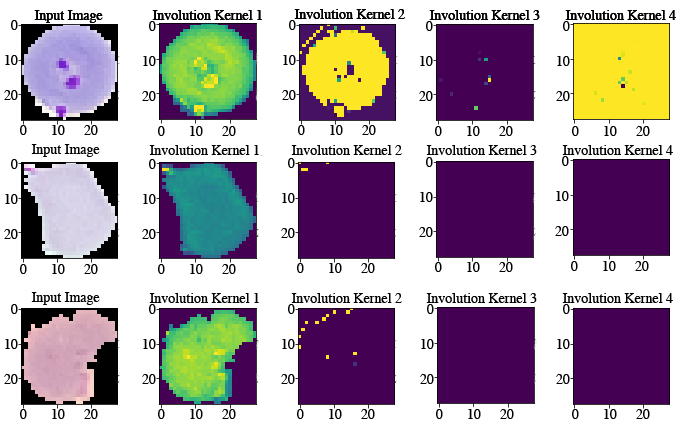}}

    \caption{Involution Kernel visualizations of Malaria Parasitized dataset}
    \label{cls_res}
\end{figure*}

Our primary goal is to show better performance by adding involution layers before the CNN-based model. Involution, with its much fewer weight parameters than traditional operations, is decisively cost-effective. Apart from the kernel, which is computed here using both the learning parameters and the pixel's value, the computation process in involution is identical to its counterpart in convolution. The process may be summarized as following Equation \ref{eq1}.


\begin{equation}
Y_{i,k,k} = \sum_{(u,u)\in \Delta_{K} }^{} H_{i,j,u + \lfloor K/2 \rfloor , v + \lfloor K/2 \rfloor , \lceil kG/C \rceil } X_{i+u,j+v,k}
 \label{eq1}
\end{equation}

The same kernel is shared across channels in involution, hence its channel-agnosticity. The actual kernel is created using the following Equation \ref{eq2}.

\begin{equation}
 H_{i,j} = \phi(X_{i,j}) = W_{1} \sigma(W_{0}X_{i,j})
 \label{eq2}
\end{equation}

The meta weights ($W_1$, $W_0$) of the kernel, meaning the weights utilized to build the kernel, are subsequently distributed across pixel values, ensuring that some shift-invariance of convolution is preserved. While it is not capable of capturing the interactions amongst the pixels as effectively, compared to attention mechanics, it is rather uniform, the calculations become faster in execution and lighter in terms of memory. Unlike attention, we do not require storing interdependencies for each pair of patch or pixel positions. Meta-weights instead suffice for their retrieval. More complex patterns than convolution can thus be constructed. Also, consequentially, the inclusion of multiple involution layers does not necessarily translate to a heavy network. 

As it reduces the model size with a reduction of convolutions, we have an improved model for medical image segmentation as well as classification in relation to memory. Involution and convolution can be construed as the equations \ref{3_eq} and \ref{4_eq}. In this context, X stands for the input data, F signifies the involution kernel, while $\sigma$ includes a non-linear activation function with data normalization. Here, H is the convolution kernel, while $\delta$ represents the activation function.

\begin{equation} 
y_{inv}=\sigma (F \ast X)
\label{3_eq}
\end{equation}

\begin{equation} 
y_{conv}=\delta (\mathcal {H} \ast y_{inv})
\label{4_eq}
\end{equation}

The output shape will automatically be aligned with the input features in the spatial dimension, to fully leverage the feature representation in the channel domain. Convolution, meanwhile, enhances possibilities of feature representation as it supplements the channel-agnostic behavior of involution.

Referring to Equation \ref{eq2}, we see that the features it generates contain shapes and their absolute position in an image. Therefore, compounding more layers does not scale to the merger of shapes as occurs in multiple layers of convolution. Instead, combining the features extracted by two layers of involution results in a four-tuple of two features and their absolute locations, whose meaningful melding would require extensive convolution. In this work, we show how the process of involution before CNN-based models can be effective. We use involution before convolutions because the involution processes an image in such a way that location-specific or spatial features are captured and assist the classification or segmentation task. Illustrations of both classification and segmentation architecture can be seen in Figure \ref{model}.

\subsection{Classification Model}
For classification, we begin with the involution block followed by the ReLU activation function and a max-pooling layer to minimize the model's weight parameters and computational costs. Then we also add batch normalization for better interlayer normalization. There are two additional convolution layers with 64 and 128 nodes that are driven by ReLU activation functions. Similar to involution, each convolution has max-pooling and batch normalization for the same purpose. Afterward, a 10\% dropout layer is added on top of the network. Finally, we employ the fully connected block, composed of four Dense layers with 256, 192, 96, and 64 nodes, to reduce all of the gathered data to a single dimension. The model concludes with a classifier layer. We refer to our proposed classification model as \textbf{Med-IC (Cls)}. Here, ``Cls'' is the shortened form of the word classification.

\subsection{Segmentation Model}
Our proposed segmentation model has six encoder blocks and five decoder blocks with three convolution layers in each block (the only exception being the last block having only a single convolution layer), with 10\% dropout applied to each blockk. The number of nodes in each block of convolutions is 16, 32, 64, 128, 256, and 512 respectively. On the other hand, for transposed convolutions in the decoder, the number of nodes is 256, 128, 64, 32, and 16. We embed one involution layer on top of the first encoder block. This follows the same idea of adding an involution layer before the main convolution-based architecture. We refer to our proposed segmentation model as \textbf{Med-IC (Seg)}. Here, ``Seg'' is the shortened form of the word segmentation. 

\section{Experiments}

\begin{table}[]
\caption{Classification performance comparison with previous works}
\begin{tabular}{c|c|c|c}
\hline
Dataset                   & Author                            & Model                                                                                      & Accuracy $\uparrow$ \\ \hline
\multirow{8}{*}{HAM10000} & Lan et al., 2022 \cite{fixcaps}   & Fixcaps                                                                                    & 96.49    \\
                          & Datta et al., 2021 \cite{soft}      & IRv2+Soft Attention                                                                        & 93.4     \\
                          & Charan et al., 2020 \cite{method}    & \begin{tabular}[c]{@{}c@{}}Loss balancing and ensemble \\ of multi-resolution\end{tabular} & 92.6     \\
                          &  Khan et al., 2024 \cite{khan2024intelligent}                                 &   Multi-deep learning models information fusion                                                                                         &   87.02        \\
                          &    Khan et al., 2021         \cite{KHAN2021106956}                      &  24-layered CNN architecture                                                                                           &  86.50          \\
                          &    Xin et al., 2022   \cite{XIN2022105939}                      &   Custom vision transformer                                                                                           &  94.30          \\
                          &       Jasil et al., 2023      \cite{jasil2023hybrid}                      &   Hybrid Densenet and residual network                                                                                           &  95.0          \\
                          &    Houssein et al., 2024         \cite{houssein2024effective}                      &   DCNN model                                                                                           &  98.50          \\
                          & Ours                              & Med-IC (Cls)                                                                                    & 98.85    \\ \hline
\multirow{8}{*}{Malaria}  &  Madhu et al., 2022 \cite{madhu2022dscn}      & DSCN-net (Capsule network with
IR)                                                                                 &  98.89    \\
                          & Kumar et al., 2024 \cite{kumar2024application}      & CNN + Capsule Network                                                                                & 99.08    \\
                          & Asif et al., 2024 \cite{asif2024malaria}      &    DBEL                                                                                 & 98.50    \\
                          & Liang et al., 2016 \cite{1111}      & Customized CNN                                                                             & 94.00    \\
                          & Rajaraman et al., 2018 \cite{rajaraman} & Designed CNN                                                                               & 94.61    \\
                          &    Elangovan and Nath, 2021 \cite{elangovan2021novel}                               &      Shallow ConvNet-18                                                                                      &   97.8       \\
                          &            Marques et al., 2022 \cite{marques2022ensemble}                       &          EfficientNetB0                                                                                  &    98.29      \\
                          &      Quan et al., 2020 \cite{quan2020effective}                             &               ADCN                                                                             &    97.47      \\
                          &    Murmu and Kumar, 2024 \cite{murmu2024dlrfnet}                             &          DLRFNet (CNN + random forest)                                                                             &    94.32      \\
                          & Ours                              & Med-IC (Cls)                                                                                       & 98.67    \\ \hline
\end{tabular}
\label{old_comp_cls}
\end{table}

\begin{table}[]
\caption{Segmentation performance comparison with previous works}
\begin{tabular}{c|c|c|c}
\hline
Dataset                 & Author and Year                                                           & Model                  & IOU $\uparrow$   \\ \hline
\multirow{8}{*}{KVASIR} & Vezakis et al., 2024 \cite{Vezakis2024-yg}        & EffiSegNet-B5          & 0.9065 \\
                        & Vezakis et al., 2024 \cite{Vezakis2024-yg}        & EffiSegNet-B4          & 0.9056 \\
                        & Dumitru et al., 2023 \cite{Dumitru2023-gc}       & DUCK-Net               & 0.9051 \\
                        & Zhou, 2023 \cite{Zhou2023-tz}                    & SEP                    & 0.9002 \\
                        & Kato et al., 2022 \cite{Kato2022-wz}             & FCBFormer             & 0.8974 \\
                        &  Biswas, 2023 \cite{biswas2023polypsamtextguidedsam}             & Polyp-SAM++             & 0.862 \\                        & Lou et al., 2023 \cite{lou2023caranet}             & CaraNet             & 0.865 \\                        &  Fan et al., 2020 \cite{10.1007/978-3-030-59725-2_26}             & PraNet             & 0.849 \\              &  Trinh et al., 2024          \cite{trinh2024samegsegmentmodelegde}             & SAM-EG             & 0.862 \\
                        
                        & Fitzgerald et al., 2024 \cite{Fitzgerald2024-rk} & FCB-SwinV2 Transformer & 0.8973 \\
                        & Srivastava et al., 2022 \cite{Srivastava2022-do} & MSRF-Net               & 0.8914 \\
                        & Ours                                                              & Med-IC (Seg)                & 0.9051 \\ \hline
\multirow{9}{*}{BUSI}  & Zhou et al., 2018  \cite{Zhou2018-cm}                               & UNet++                 & 0.6433 \\
                        & Zhang et al., 2018 \cite{Zhang2018-go}                              & ResUNet                & 0.6489 \\
                        & Valanarasu et al., 2021\cite{Valanarasu2021-ig}                         & MedT                   & 0.6389 \\
                        & Chen et al., 2021 \cite{Chen2021-tp}                               & TransUNet              & 0.6692 \\
                        & Valanarasu et al., 2022 \cite{Valanarasu2022-nz}                         & UNeXt                  & 0.6695 \\
                        & Xu et al., 2023 \cite{10016712}                                  & RMTL-Net               & 0.7193 \\
                        & Jin et al., 2023 \cite{jin2023novel}                              & MBSNet                 & 0.6321 \\
                        & Jiao et al., 2024 \cite{JIAO2024103202}                            & USFM                   & 0.7600 \\
                        & Bobowicz et al., 2024 \cite{BOBOWICZ2024105522}                            & PraNet                   & 0.73 \\
                        & Bobowicz et al., 2024 \cite{BOBOWICZ2024105522}                            & CaraNet                   & 0.73 \\
                        & Bobowicz et al., 2024 \cite{BOBOWICZ2024105522}                            & FCBFormer                   & 0.80 \\
                        & Ours                                                              & Med-IC (Seg)                 & 0.8262 \\ \hline
\end{tabular}
\label{old_comp_seg}
\end{table}

\subsection{Dataset}

We use four different datasets in total. For the classification task, the HAM10000 dataset \cite{ham} and Malaria Screener \cite{doiMalariaScreener}. We randomly split both datasets into training sets of 80\% and testing sets of 20\% for every class. Additionally, we split the arbitrarily selected training dataset into 90\% for training and 10\% for validation. The image size for classification is processed to $28 \times 28$. For segmentation, we use the Kvasir-SEG dataset \cite{jha2020kvasir} and a public dataset of breast ultrasound images (BUSI) \cite{ALDHABYANI2020104863}, of women between the ages of 25 and 75. The image size for segmentation tasks is $128 \times 128$.

\subsection{Experimental Setup and Details}
For this experiment, the models are trained on an NVIDIA GeForce RTX 3080Ti GPU with a performance of 34.1 TeraFLOPS.

For the ablation study, we add more involution layers in Med-IC (Cls) creating three hybrid variations. Keeping the architectural designs identical to the proposed model, we also utilize a one-layer involution neural network (INN) and a three-layer convolutional neural network (CNN) with 256 nodes in the last convolutional layer. For HAM10000 and Malaria, we train the model with 30 epochs, a batch size of 16, and the Adam optimizer with a 0.0001 learning rate which has given both the best fit and overall performance. 

In segmentation, we utilize five different models which are identical to our proposed model. The U-Net (without extra convolutions) model has one convolution layer with 512 nodes in the 6th encoder block. The heavier variant has two extra convolution layers with 512 nodes making the number of model parameters more than 11.5 million. On the other hand, we utilize the three involution-embedded model to establish that the multiple involution layer is unnecessary for this task, with one, two, and three involution layers embedded on top of the encoder blocks of the U-Net. 
The dropout rate is kept the same as the classification models. For segmentation, we train the models with 100 epochs, a batch size of 8, and the Adam optimizer with a 0.00001 learning rate which results in the best outcome. 

For classification, we employ accuracy, recall or sensitivity, and F1-score for our evaluation metrics.  In segmentation, we use accuracy, intersection over union (IoU), and dice coefficient (DSC). For a better understanding of the predictions (in ablation), we use Grad-CAM \cite{Selvaraju_2019} for segmentation and involution kernel visualizations for classification.

\subsection{Ablation Studies}

From Figure \ref{res}, we can see the predictions where the Med-IC (Seg) with only one involution layer gives the best prediction mask overall. The closest one (U-Net with extra convolution layers) in terms of prediction has almost double the number of weight parameters and fails to capture features adequately with slightly poorer metrics. In involution-embedded networks, a single involution layer-based model performs the best empirically. With the increase of involution layers, however, the performance of the model starts to decline. As per the Grad-CAM visualization and sizes of the prediction masks, the addition of further involution layers assimilates insubstantial features, which leads to inaccurate segmentation. The prediction mask, in particular, disperses throughout the image and becomes highly non-localized. The effect of multiple involutions can be seen clearly in Figure \ref{cls_res}. We can see how the information is getting lost due to the strength of location-specific extraction. The images are similar due to having only one particular anomaly or target region, making the use of multiple involutions in the hybrid model negatively effective, which is mentioned in Table \ref{abla_cls} for classification and Table \ref{abla_seg} for segmentation.

\begin{table}[]
\centering
\caption{Segmentation performance comparison of different variations of the hybrid models}
\begin{tabular}{c|c|ccc|ccc}
\hline
\multirow{2}{*}{Model}                                                         & \multirow{2}{*}{Weight Parameters $\downarrow$} & \multicolumn{3}{c|}{KVASIR}                                  & \multicolumn{3}{c}{BUSI}                           \\ \cline{3-8} 
                                                                               &                                       & \multicolumn{1}{c|}{Accuracy $\uparrow$} & \multicolumn{1}{c|}{IOU $\uparrow$}   & DSC $\uparrow$ & \multicolumn{1}{c|}{Accuracy $\uparrow$} & \multicolumn{1}{c|}{IOU $\uparrow$}  & DSC $\uparrow$ \\ \hline
\begin{tabular}[c]{@{}c@{}}Vanilla U-Net\\ (Without extra layers)\end{tabular} & 6,988,113                             & \multicolumn{1}{c|}{83.62}    & \multicolumn{1}{c|}{0.7862}  &  0.8099   & \multicolumn{1}{c|}{85.33}    & \multicolumn{1}{c|}{0.6104} &   0.6923  \\
\begin{tabular}[c]{@{}c@{}}Vanilla U-Net\\ (With extra layers)\end{tabular}    & 11,707,729                            & \multicolumn{1}{c|}{89.22}    & \multicolumn{1}{c|}{0.8214}  &  0.8489   & \multicolumn{1}{c|}{89.22}    & \multicolumn{1}{c|}{0.6433} &  0.7190   \\
\begin{tabular}[c]{@{}c@{}} Hybrid-1 or Med-IC (Seg) \end{tabular}      & 6,988,139                             & \multicolumn{1}{c|}{93.89}    & \multicolumn{1}{c|}{0.9051} &  0.9463   & \multicolumn{1}{c|}{93.75}    & \multicolumn{1}{c|}{0.8162} &  0.8550   \\
\begin{tabular}[c]{@{}c@{}} Hybrid-2 \end{tabular}     & 6,988,165                             & \multicolumn{1}{c|}{89.76}    & \multicolumn{1}{c|}{0.8550}  &  0.8800   & \multicolumn{1}{c|}{92.15}    & \multicolumn{1}{c|}{0.7753} &   0.7908  \\
\begin{tabular}[c]{@{}c@{}} Hybrid-3  \end{tabular}     & 6,988,191                             & \multicolumn{1}{c|}{87.12}    & \multicolumn{1}{c|}{0.7867}  &  0.8232   & \multicolumn{1}{c|}{84.28}    & \multicolumn{1}{c|}{0.6089} &   0.6667  \\ \hline
\end{tabular}
\label{abla_seg}
\end{table}

\subsection{Results and Comparison}

\begin{table}[]
\centering
\caption{Performance comparison with other models.}
\begin{tabular}{cccccc}

\hline
\multicolumn{6}{c}{Classification}                                                                                                   \\ \hline
\multicolumn{1}{c|}{\multirow{2}{*}{Model}} & \multicolumn{1}{c|}{\multirow{2}{*}{Weight Parameters $\downarrow$ }} & \multicolumn{2}{c|}{HAM10000}                               & \multicolumn{2}{c}{Malaria}       \\ \cline{3-6} 
\multicolumn{1}{c|}{}                       & \multicolumn{1}{c|}{}                                   & \multicolumn{1}{c|}{Accuracy $\uparrow$} & \multicolumn{1}{c|}{Recall $\uparrow$ } & \multicolumn{1}{c|}{Accuracy $\uparrow$} & Recall  $\uparrow$  \\ \hline
\multicolumn{1}{c|}{Vanilla CNN (4 Layers)}                          & \multicolumn{1}{c|}{6.7 Million}                     & \multicolumn{1}{c|}{95.14}         & \multicolumn{1}{c|}{93.98}       & \multicolumn{1}{c|}{97.14}    &    96.67    \\
\multicolumn{1}{c|}{ResNet50}                       & \multicolumn{1}{c|}{23.71 Million}                                   & \multicolumn{1}{c|}{93.23}         & \multicolumn{1}{c|}{93.10}       & \multicolumn{1}{c|}{95.55}    &     95.00   \\
\multicolumn{1}{c|}{DenseNet201}                       & \multicolumn{1}{c|}{18.37 Million}                                   & \multicolumn{1}{c|}{91.73}         & \multicolumn{1}{c|}{90.50}       & \multicolumn{1}{c|}{93.67}    &    92.44    \\
\multicolumn{1}{c|}{NasNetMobile}                       & \multicolumn{1}{c|}{4.33 Million}                                   & \multicolumn{1}{c|}{90.33}         & \multicolumn{1}{c|}{89.99}       & \multicolumn{1}{c|}{91.31}    &   90.74     \\
\multicolumn{1}{c|}{VGG19}                       & \multicolumn{1}{c|}{25.43 Million}                                   & \multicolumn{1}{c|}{92.22}         & \multicolumn{1}{c|}{91.25}       & \multicolumn{1}{c|}{94.14}    &    93.69    \\
\multicolumn{1}{c|}{Xception}                       & \multicolumn{1}{c|}{36.05 Million}                                   & \multicolumn{1}{c|}{93.84}         & \multicolumn{1}{c|}{93.04}       & \multicolumn{1}{c|}{94.44}    &    94.50    \\
\multicolumn{1}{c|}{Vision Transformer (Vanilla)}                       & \multicolumn{1}{c|}{26 Million}                                   & \multicolumn{1}{c|}{94.29}         & \multicolumn{1}{c|}{93.02}       & \multicolumn{1}{c|}{95.91}    &    94.39    \\
\multicolumn{1}{c|}{ConvNextTiny}                       & \multicolumn{1}{c|}{34.7 Million }                                   & \multicolumn{1}{c|}{93.50}         & \multicolumn{1}{c|}{92.34}       & \multicolumn{1}{c|}{95.83}    &    94.77     \\ 
\multicolumn{1}{c|}{Ours}                       & \multicolumn{1}{c|}{2.11 Million}                                   & \multicolumn{1}{c|}{98.85}         & \multicolumn{1}{c|}{99.41}       & \multicolumn{1}{c|}{98.67}    &    98.55    \\ \hline
\multicolumn{6}{c}{Segmentation}                                                                                                                                                                        \\ \hline
\multicolumn{1}{c|}{\multirow{2}{*}{Model}} & \multicolumn{1}{c|}{\multirow{2}{*}{Weight Parameters $\downarrow$}} & \multicolumn{2}{c|}{KVASIR}                                 & \multicolumn{2}{c}{BUSI}          \\ \cline{3-6} 
\multicolumn{1}{c|}{}                       & \multicolumn{1}{c|}{}                                   & \multicolumn{1}{c|}{IoU $\uparrow$}      & \multicolumn{1}{c|}{DSC $\uparrow$}    & \multicolumn{1}{c|}{IoU $\uparrow$} & DSC  $\uparrow$  \\ \hline
\multicolumn{1}{c|}{Vanilla U-Net (With extra convolutions)}                       & \multicolumn{1}{c|}{11.7 Million}                                   & \multicolumn{1}{c|}{0.8214}         & \multicolumn{1}{c|}{0.8489}       & \multicolumn{1}{c|}{0.6433}    &    0.719    \\
\multicolumn{1}{c|}{Attention U-Net}                       & \multicolumn{1}{c|}{7.09 Million}                                   & \multicolumn{1}{c|}{0.8334}         & \multicolumn{1}{c|}{0.862}       & \multicolumn{1}{c|}{0.7393}    &  0.8010      \\ 
\multicolumn{1}{c|}{Ours}                       & \multicolumn{1}{c|}{6.98 Million}                                   & \multicolumn{1}{c|}{0.9051}         & \multicolumn{1}{c|}{0.9463}       & \multicolumn{1}{c|}{0.8162}    &    0.855    \\ \hline
\end{tabular}
\label{model_comp}
\end{table}





\begin{figure*}
  \centering
  \setkeys{Gin}{width=1\linewidth}
\begin{subfigure}{15cm}  
 \includegraphics{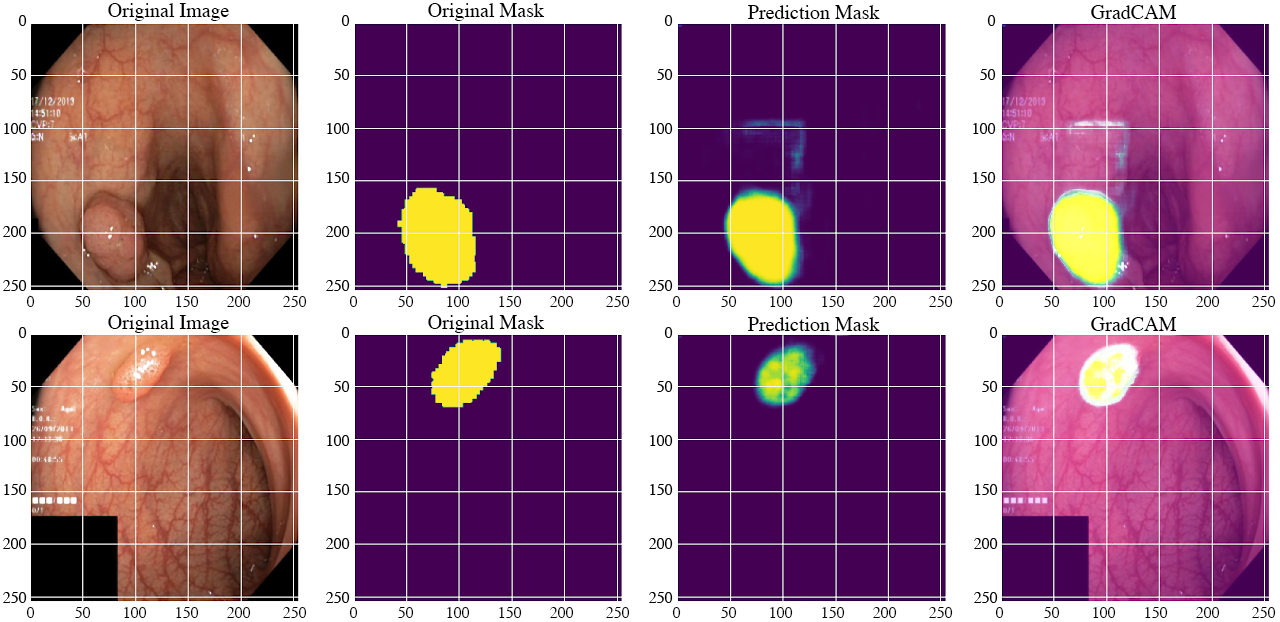}
 \caption{Sample predictions from KVASIR dataset.}
 \label{fig1a}
\end{subfigure}

\begin{subfigure}{15cm}  
 \includegraphics{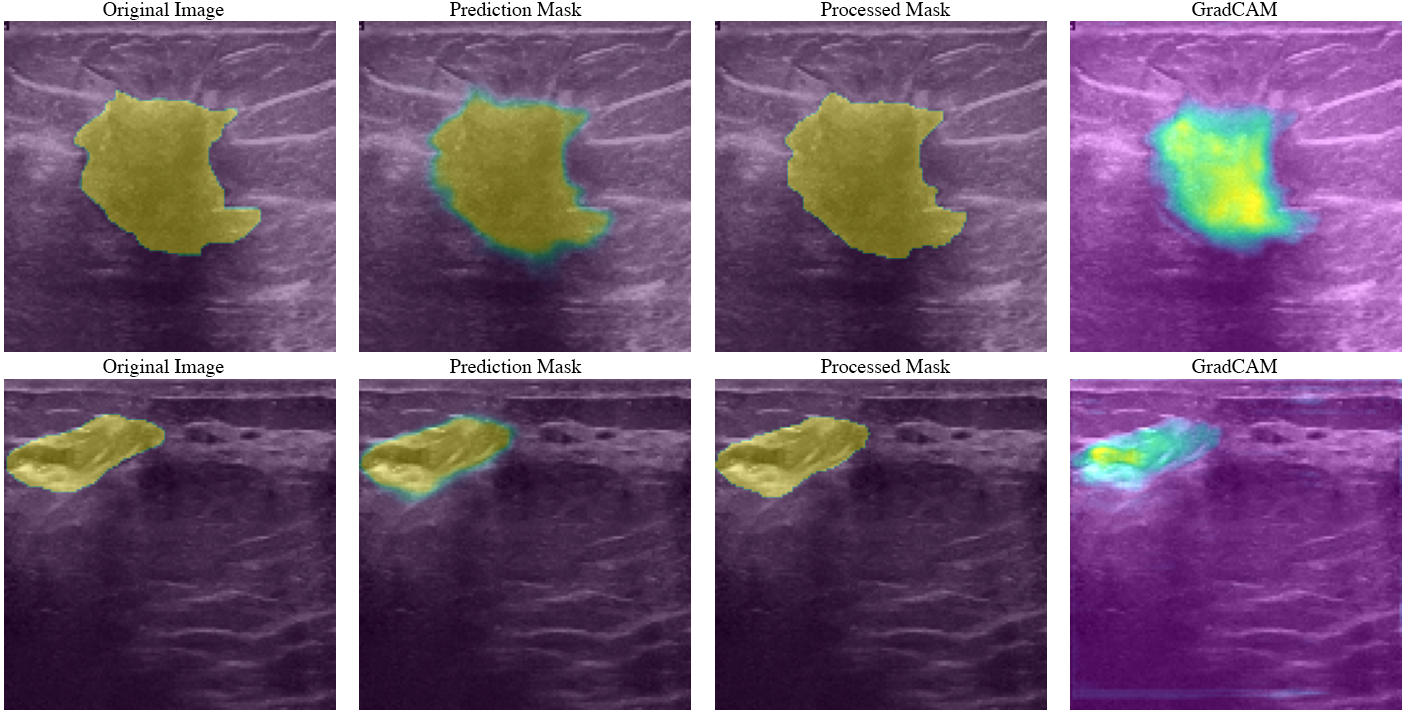}
 \caption{Sample predictions from BUSI dataset.}
 \label{fig1b}
\end{subfigure}
\caption{Sample predictions of our proposed model.}
\label{fig2}
\end{figure*}

In the classification task, we employ ResNet50, DenseNet201 NasNetMobile, VGG19, Xception, Vision Transformer, and ConvNextTiny using the same Adam optimizer with a 0.0001 learning rate for comparison. The batch size is 16 in classification similar to Med-IC (Cls). For segmentation, we compare our proposed model with Vanilla U-Net and Attention U-Net. In this scenario, the hyperparameters are the same due to getting optimal performance with the same values in these models as well. Figure \ref{model_comp} shows the full comparison, where we can see that our proposed Med-IC (Cls) and Med-IC (Seg) performs the best.

Coming to the comparison with existing literature, for classification we get the full comparison in Table \ref{old_comp_cls}. Med-IC (Cls) outperforms all in HAM10000. Our proposed model also performs dominantly in the malaria dataset but capsule network-based models \cite{madhu2022dscn, kumar2024application} are at the top of the list. However, the good performance here indicates better generalizability and proper utilization of location-specific features that both HAM10000 and malaria dataset images hold. Table \ref{old_comp_seg} displays the overall comparison in segmentation. Similarly, the concept of adding a single layer involution to enhance performance proves right. Our proposed model outperforms all models in both cases. The only exception is EffiSegNet variants \cite{Vezakis2024-yg}. EffiSegNet-B5 and EffiSegNet-B6 have 28.3 and 40.7 million weight parameters respectively, whereas our model possesses only 6.98 million parameters. In the same study, the lighter variants (6-7 million parameters) have achieved less than 0.89 IoU.

The segmentation outputs can be seen in Figure \ref{fig2}. The KVASIR prediction plots are in Figure \ref{fig1a} and the BUSI prediction plots are in Figure \ref{fig1b}. In both cases, we see that our proposed model accurately and precisely segments the anomaly region.



\section{Discussion}


The visual divergence in patterns at different locations within an image can be challenging to learn in medical image recognition tasks.  Even though involution has shown promise as a feature extractor in several studies, our review suggests a gap in addressing this specific challenge directly. Classification and segmentation tasks in the aforementioned problem of medical image identification, where little detail is vital for clinical purposes, are greatly enhanced with the location-specific strategy of involution. Additionally, this task's size efficiency has been greatly increased, resulting in a notably lower computational cost. This is because computational cost is very crucial for deep learning tasks in this era of artificial intelligence \cite{greenai}. There are significantly fewer weight parameters in involution than in convolution. In summary, the problem of effectively extracting spatial information across diverse regions while maintaining resource efficiency within the model's framework is a significant yet unexplored area in deep learning and medical imaging.


\section{Conclusion and Future Work}

Our research explores the integration of an involution layer prior to a convolution-based model. Multiple involution layers in cell-like images have been shown to reduce efficacy due to spatial details' over-extraction. To address this, our proposed approach strategically incorporates involution, which adds minimal weight parameters, and yet yields superior performance compared to a single convolution operation, which involves a significantly higher number of weight parameters. Our results across four diverse datasets consistently demonstrate this pattern, suggesting that this approach can significantly enhance resource efficiency in medical diagnostic procedures. We also believe this technique holds promising implications for generative vision models, showcasing its potential for widespread impact.

\balance
\bibliographystyle{elsarticle-num}
\bibliography{main}

\begin{thebibliography}{10}
\expandafter\ifx\csname url\endcsname\relax
  \def\url#1{\texttt{#1}}\fi
\expandafter\ifx\csname urlprefix\endcsname\relax\def\urlprefix{URL }\fi
\expandafter\ifx\csname href\endcsname\relax
  \def\href#1#2{#2} \def\path#1{#1}\fi

\bibitem{JIANG2023106726}
H.~Jiang, Z.~Diao, T.~Shi, Y.~Zhou, F.~Wang, W.~Hu, X.~Zhu, S.~Luo, G.~Tong, Y.-D. Yao, \href{https://www.sciencedirect.com/science/article/pii/S0010482523001919}{A review of deep learning-based multiple-lesion recognition from medical images: classification, detection and segmentation}, Computers in Biology and Medicine 157 (2023) 106726.
\newblock \href {https://doi.org/https://doi.org/10.1016/j.compbiomed.2023.106726} {\path{doi:https://doi.org/10.1016/j.compbiomed.2023.106726}}.
\newline\urlprefix\url{https://www.sciencedirect.com/science/article/pii/S0010482523001919}

\bibitem{alzubaidi_2021_review}
L.~Alzubaidi, J.~Zhang, A.~J. Humaidi, A.~Al-Dujaili, Y.~Duan, O.~Al-Shamma, J.~Santamaría, M.~A. Fadhel, M.~Al-Amidie, L.~Farhan, \href{https://journalofbigdata.springeropen.com/articles/10.1186/s40537-021-00444-8}{Review of deep learning: concepts, cnn architectures, challenges, applications, future directions}, Journal of Big Data 8 (03 2021).
\newblock \href {https://doi.org/10.1186/s40537-021-00444-8} {\path{doi:10.1186/s40537-021-00444-8}}.
\newline\urlprefix\url{https://journalofbigdata.springeropen.com/articles/10.1186/s40537-021-00444-8}

\bibitem{pr1}
J.~Lee, S.~Narang, J.~J. Martinez, G.~Rao, A.~U.~K. Rao, \href{https://doi.org/10.1117/1.JMI.2.4.041006}{{Associating spatial diversity features of radiologically defined tumor habitats with epidermal growth factor receptor driver status and 12-month survival in glioblastoma: methods and preliminary investigation}}, Journal of Medical Imaging 2~(4) (2015) 041006.
\newblock \href {https://doi.org/10.1117/1.JMI.2.4.041006} {\path{doi:10.1117/1.JMI.2.4.041006}}.
\newline\urlprefix\url{https://doi.org/10.1117/1.JMI.2.4.041006}

\bibitem{pr2}
S.~Leng, L.~Yu, L.~Chen, J.~C.~R. Giraldo, C.~H. McCollough, \href{https://doi.org/10.1117/12.912126}{{Correlation between model observer and human observer performance in CT imaging when lesion location is uncertain}}, in: N.~J. Pelc, R.~M. Nishikawa, B.~R. Whiting (Eds.), Medical Imaging 2012: Physics of Medical Imaging, Vol. 8313, International Society for Optics and Photonics, SPIE, 2012, p. 83131M.
\newblock \href {https://doi.org/10.1117/12.912126} {\path{doi:10.1117/12.912126}}.
\newline\urlprefix\url{https://doi.org/10.1117/12.912126}

\bibitem{pr3}
D.~R. Sarvamangala, R.~V. Kulkarni, Convolutional neural networks in medical image understanding: a survey, Evol. Intell. 15~(1) (2022) 1--22.

\bibitem{HOSSEINI2024100357}
M.~S. Hosseini, B.~E. Bejnordi, V.~Q.-H. Trinh, L.~Chan, D.~Hasan, X.~Li, S.~Yang, T.~Kim, H.~Zhang, T.~Wu, K.~Chinniah, S.~Maghsoudlou, R.~Zhang, J.~Zhu, S.~Khaki, A.~Buin, F.~Chaji, A.~Salehi, B.~N. Nguyen, D.~Samaras, K.~N. Plataniotis, \href{https://www.sciencedirect.com/science/article/pii/S2153353923001712}{Computational pathology: A survey review and the way forward}, Journal of Pathology Informatics 15 (2024) 100357.
\newblock \href {https://doi.org/https://doi.org/10.1016/j.jpi.2023.100357} {\path{doi:https://doi.org/10.1016/j.jpi.2023.100357}}.
\newline\urlprefix\url{https://www.sciencedirect.com/science/article/pii/S2153353923001712}

\bibitem{JAVED2021102104}
S.~Javed, A.~Mahmood, J.~Dias, N.~Werghi, N.~Rajpoot, \href{https://www.sciencedirect.com/science/article/pii/S136184152100150X}{Spatially constrained context-aware hierarchical deep correlation filters for nucleus detection in histology images}, Medical Image Analysis 72 (2021) 102104.
\newblock \href {https://doi.org/https://doi.org/10.1016/j.media.2021.102104} {\path{doi:https://doi.org/10.1016/j.media.2021.102104}}.
\newline\urlprefix\url{https://www.sciencedirect.com/science/article/pii/S136184152100150X}

\bibitem{involution_paper}
D.~Li, J.~Hu, C.~Wang, X.~Li, Q.~She, L.~Zhu, T.~Zhang, Q.~Chen, \href{https://doi.ieeecomputersociety.org/10.1109/CVPR46437.2021.01214}{Involution: Inverting the inherence of convolution for visual recognition}, in: 2021 IEEE/CVF Conference on Computer Vision and Pattern Recognition (CVPR), IEEE Computer Society, Los Alamitos, CA, USA, 2021, pp. 12316--12325.
\newblock \href {https://doi.org/10.1109/CVPR46437.2021.01214} {\path{doi:10.1109/CVPR46437.2021.01214}}.
\newline\urlprefix\url{https://doi.ieeecomputersociety.org/10.1109/CVPR46437.2021.01214}

\bibitem{derm6}
H.~Younis, M.~H. Bhatti, M.~Azeem, Classification of skin cancer dermoscopy images using transfer learning, 2019 15th International Conference on Emerging Technologies (ICET) (2019) 1--4.

\bibitem{id12}
S.~Rajaraman, S.~K. Antani, M.~Poostchi, K.~Silamut, M.~A. Hossain, R.~J. Maude, S.~Jaeger, G.~R. Thoma, Pre-trained convolutional neural networks as feature extractors toward improved malaria parasite detection in thin blood smear images, PeerJ 6 (2018) e4568.

\bibitem{id39}
K.~He, X.~Zhang, S.~Ren, J.~Sun, Deep residual learning for image recognition, in: 2016 IEEE Conference on Computer Vision and Pattern Recognition (CVPR), 2016, pp. 770--778.
\newblock \href {https://doi.org/10.1109/CVPR.2016.90} {\path{doi:10.1109/CVPR.2016.90}}.

\bibitem{id40}
G.~Huang, Z.~Liu, L.~Van Der~Maaten, K.~Q. Weinberger, Densely connected convolutional networks, in: 2017 IEEE Conference on Computer Vision and Pattern Recognition (CVPR), 2017, pp. 2261--2269.
\newblock \href {https://doi.org/10.1109/CVPR.2017.243} {\path{doi:10.1109/CVPR.2017.243}}.

\bibitem{id43}
F.~Wang, M.~Jiang, C.~Qian, S.~Yang, C.~Li, H.~Zhang, X.~Wang, X.~Tang, Residual attention network for image classification, in: 2017 IEEE Conference on Computer Vision and Pattern Recognition (CVPR), 2017, pp. 6450--6458.
\newblock \href {https://doi.org/10.1109/CVPR.2017.683} {\path{doi:10.1109/CVPR.2017.683}}.

\bibitem{ham}
N.~C.~F. Codella, V.~Rotemberg, P.~Tschandl, M.~E. Celebi, S.~W. Dusza, D.~A. Gutman, B.~Helba, A.~Kalloo, K.~Liopyris, M.~A. Marchetti, H.~Kittler, A.~Halpern, \href{http://arxiv.org/abs/1902.03368}{Skin lesion analysis toward melanoma detection 2018: {A} challenge hosted by the international skin imaging collaboration {(ISIC)}}, CoRR abs/1902.03368 (2019).
\newblock \href {http://arxiv.org/abs/1902.03368} {\path{arXiv:1902.03368}}.
\newline\urlprefix\url{http://arxiv.org/abs/1902.03368}

\bibitem{CASSIDY2022102305}
B.~Cassidy, C.~Kendrick, A.~Brodzicki, J.~Jaworek-Korjakowska, M.~H. Yap, \href{https://www.sciencedirect.com/science/article/pii/S1361841521003509}{Analysis of the isic image datasets: Usage, benchmarks and recommendations}, Medical Image Analysis 75 (2022) 102305.
\newblock \href {https://doi.org/https://doi.org/10.1016/j.media.2021.102305} {\path{doi:https://doi.org/10.1016/j.media.2021.102305}}.
\newline\urlprefix\url{https://www.sciencedirect.com/science/article/pii/S1361841521003509}

\bibitem{9446143}
N.~Siddique, S.~Paheding, C.~P. Elkin, V.~Devabhaktuni, U-net and its variants for medical image segmentation: A review of theory and applications, IEEE Access 9 (2021) 82031--82057.
\newblock \href {https://doi.org/10.1109/ACCESS.2021.3086020} {\path{doi:10.1109/ACCESS.2021.3086020}}.

\bibitem{10643318}
R.~Azad, E.~K. Aghdam, A.~Rauland, Y.~Jia, A.~H. Avval, A.~Bozorgpour, S.~Karimijafarbigloo, J.~P. Cohen, E.~Adeli, D.~Merhof, Medical image segmentation review: The success of u-net, IEEE Transactions on Pattern Analysis and Machine Intelligence (2024) 1--20\href {https://doi.org/10.1109/TPAMI.2024.3435571} {\path{doi:10.1109/TPAMI.2024.3435571}}.

\bibitem{BOBOWICZ2024105522}
M.~Bobowicz, M.~Badocha, K.~Gwozdziewicz, M.~Rygusik, P.~Kalinowska, E.~Szurowska, T.~Dziubich, \href{https://www.sciencedirect.com/science/article/pii/S1386505624001850}{Segmentation-based bi-rads ensemble classification of breast tumours in ultrasound images}, International Journal of Medical Informatics 189 (2024) 105522.
\newblock \href {https://doi.org/https://doi.org/10.1016/j.ijmedinf.2024.105522} {\path{doi:https://doi.org/10.1016/j.ijmedinf.2024.105522}}.
\newline\urlprefix\url{https://www.sciencedirect.com/science/article/pii/S1386505624001850}

\bibitem{SARICA2023104965}
B.~Sarica, D.~Z. Seker, B.~Bayram, \href{https://www.sciencedirect.com/science/article/pii/S1386505622002799}{A dense residual u-net for multiple sclerosis lesions segmentation from multi-sequence 3d mr images}, International Journal of Medical Informatics 170 (2023) 104965.
\newblock \href {https://doi.org/https://doi.org/10.1016/j.ijmedinf.2022.104965} {\path{doi:https://doi.org/10.1016/j.ijmedinf.2022.104965}}.
\newline\urlprefix\url{https://www.sciencedirect.com/science/article/pii/S1386505622002799}

\bibitem{NIDA201937}
N.~Nida, A.~Irtaza, A.~Javed, M.~H. Yousaf, M.~T. Mahmood, \href{https://www.sciencedirect.com/science/article/pii/S1386505618307470}{Melanoma lesion detection and segmentation using deep region based convolutional neural network and fuzzy c-means clustering}, International Journal of Medical Informatics 124 (2019) 37--48.
\newblock \href {https://doi.org/https://doi.org/10.1016/j.ijmedinf.2019.01.005} {\path{doi:https://doi.org/10.1016/j.ijmedinf.2019.01.005}}.
\newline\urlprefix\url{https://www.sciencedirect.com/science/article/pii/S1386505618307470}

\bibitem{GUO2019105}
S.~Guo, K.~Wang, H.~Kang, Y.~Zhang, Y.~Gao, T.~Li, \href{https://www.sciencedirect.com/science/article/pii/S138650561831195X}{Bts-dsn: Deeply supervised neural network with short connections for retinal vessel segmentation}, International Journal of Medical Informatics 126 (2019) 105--113.
\newblock \href {https://doi.org/https://doi.org/10.1016/j.ijmedinf.2019.03.015} {\path{doi:https://doi.org/10.1016/j.ijmedinf.2019.03.015}}.
\newline\urlprefix\url{https://www.sciencedirect.com/science/article/pii/S138650561831195X}

\bibitem{Rahman_2023_WACV}
M.~M. Rahman, R.~Marculescu, Medical image segmentation via cascaded attention decoding, in: Proceedings of the IEEE/CVF Winter Conference on Applications of Computer Vision (WACV), 2023, pp. 6222--6231.

\bibitem{10115109}
M.~F. Islam, S.~Zabeen, F.~B. Rahman, M.~A. Islam, F.~B. Kibria, M.~A. Manab, D.~Z. Karim, A.~A. Rasel, Unic-net: Uncertainty aware involution-convolution hybrid network for two-level disease identification, in: SoutheastCon 2023, 2023, pp. 305--312.
\newblock \href {https://doi.org/10.1109/SoutheastCon51012.2023.10115109} {\path{doi:10.1109/SoutheastCon51012.2023.10115109}}.

\bibitem{islam2024involution}
M.~F. Islam, M.~A. Manab, J.~J. Mondal, S.~Zabeen, F.~B. Rahman, M.~Z. Hasan, F.~Sadeque, J.~Noor, Involution fused convnet for classifying eye-tracking patterns of children with autism spectrum disorder (2024).
\newblock \href {http://arxiv.org/abs/2401.03575} {\path{arXiv:2401.03575}}.

\bibitem{ICNet}
G.~Liang, H.~Wang, I-cnet: Leveraging involution and convolution for image classification, IEEE Access 10 (2022) 2077--2082.
\newblock \href {https://doi.org/10.1109/ACCESS.2021.3139464} {\path{doi:10.1109/ACCESS.2021.3139464}}.

\bibitem{10288488}
A.~A. Asiri, A.~Shaf, T.~Ali, M.~Zafar, M.~A. Pasha, M.~Irfan, S.~Alqahtani, A.~J. Alghamdi, A.~H. Alghamdi, A.~F.~A. Alshamrani, M.~Aleylyani, S.~Alamri, Enhancing brain tumor diagnosis: Transitioning from convolutional neural network to involutional neural network, IEEE Access 11 (2023) 123080--123095.
\newblock \href {https://doi.org/10.1109/ACCESS.2023.3326421} {\path{doi:10.1109/ACCESS.2023.3326421}}.

\bibitem{TIAN2022487}
Y.~Tian, X.~Gong, J.~Tang, B.~Su, X.~Liu, X.~Zhang, \href{https://www.sciencedirect.com/science/article/pii/S0893608022001885}{Giu-gans: Global information utilization for generative adversarial networks}, Neural Networks 152 (2022) 487--498.
\newblock \href {https://doi.org/https://doi.org/10.1016/j.neunet.2022.05.014} {\path{doi:https://doi.org/10.1016/j.neunet.2022.05.014}}.
\newline\urlprefix\url{https://www.sciencedirect.com/science/article/pii/S0893608022001885}

\bibitem{9948278}
H.~Xiao, L.~Peng, S.~Peng, Y.~Zhang, Lung image segmentation based on involution unet model, in: 2022 5th International Conference on Advanced Electronic Materials, Computers and Software Engineering (AEMCSE), 2022, pp. 184--187.
\newblock \href {https://doi.org/10.1109/AEMCSE55572.2022.00045} {\path{doi:10.1109/AEMCSE55572.2022.00045}}.

\bibitem{9665787}
G.~Liang, H.~Wang, I-cnet: Leveraging involution and convolution for image classification, IEEE Access 10 (2022) 2077--2082.
\newblock \href {https://doi.org/10.1109/ACCESS.2021.3139464} {\path{doi:10.1109/ACCESS.2021.3139464}}.

\bibitem{10112364}
J.~Nidamanuri, S.~Vaigarai, V.~S. Gunavardhan, K.~V. Teja, Sced: A multi-criteria-based modeling of deep learning architectures for skin cancer analysis, in: 2023 10th International Conference on Computing for Sustainable Global Development (INDIACom), 2023, pp. 1110--1115.

\bibitem{ICObj}
T.~Gao, Y.~Li, X.~Han, A convolution-involution hybrid framework for monocular 3d object detection, in: 2021 IEEE 4th International Conference on Computer and Communication Engineering Technology (CCET), 2021, pp. 41--47.
\newblock \href {https://doi.org/10.1109/CCET52649.2021.9544225} {\path{doi:10.1109/CCET52649.2021.9544225}}.

\bibitem{fixcaps}
Z.~Lan, S.~Cai, X.~He, X.~Wen, Fixcaps: An improved capsules network for diagnosis of skin cancer, IEEE Access 10 (2022) 76261--76267.
\newblock \href {https://doi.org/10.1109/ACCESS.2022.3181225} {\path{doi:10.1109/ACCESS.2022.3181225}}.

\bibitem{soft}
S.~K. Datta, M.~A. Shaikh, S.~N. Srihari, M.~Gao, Soft attention improves skin cancer classification performance, in: M.~Reyes, P.~Henriques~Abreu, J.~Cardoso, M.~Hajij, G.~Zamzmi, P.~Rahul, L.~Thakur (Eds.), Interpretability of Machine Intelligence in Medical Image Computing, and Topological Data Analysis and Its Applications for Medical Data, Springer International Publishing, Cham, 2021, pp. 13--23.

\bibitem{method}
D.~S. Charan, H.~Nadipineni, S.~Sahayam, U.~Jayaraman, \href{https://arxiv.org/abs/2008.09418}{Method to classify skin lesions using dermoscopic images} (2020).
\newblock \href {https://doi.org/10.48550/ARXIV.2008.09418} {\path{doi:10.48550/ARXIV.2008.09418}}.
\newline\urlprefix\url{https://arxiv.org/abs/2008.09418}

\bibitem{khan2024intelligent}
M.~A. Khan, K.~Muhammad, M.~Sharif, T.~Akram, S.~Kadry, Intelligent fusion-assisted skin lesion localization and classification for smart healthcare, Neural Computing and Applications 36~(1) (2024) 37--52.

\bibitem{KHAN2021106956}
M.~A. Khan, Y.-D. Zhang, M.~Sharif, T.~Akram, \href{https://www.sciencedirect.com/science/article/pii/S0045790620308028}{Pixels to classes: Intelligent learning framework for multiclass skin lesion localization and classification}, Computers \& Electrical Engineering 90 (2021) 106956.
\newblock \href {https://doi.org/https://doi.org/10.1016/j.compeleceng.2020.106956} {\path{doi:https://doi.org/10.1016/j.compeleceng.2020.106956}}.
\newline\urlprefix\url{https://www.sciencedirect.com/science/article/pii/S0045790620308028}

\bibitem{XIN2022105939}
C.~Xin, Z.~Liu, K.~Zhao, L.~Miao, Y.~Ma, X.~Zhu, Q.~Zhou, S.~Wang, L.~Li, F.~Yang, S.~Xu, H.~Chen, \href{https://www.sciencedirect.com/science/article/pii/S0010482522006746}{An improved transformer network for skin cancer classification}, Computers in Biology and Medicine 149 (2022) 105939.
\newblock \href {https://doi.org/https://doi.org/10.1016/j.compbiomed.2022.105939} {\path{doi:https://doi.org/10.1016/j.compbiomed.2022.105939}}.
\newline\urlprefix\url{https://www.sciencedirect.com/science/article/pii/S0010482522006746}

\bibitem{jasil2023hybrid}
S.~G. Jasil, V.~Ulagamuthalvi, A hybrid cnn architecture for skin lesion classification using deep learning, Soft Computing (2023) 1--10.

\bibitem{houssein2024effective}
E.~H. Houssein, D.~A. Abdelkareem, G.~Hu, M.~A. Hameed, I.~A. Ibrahim, M.~Younan, An effective multiclass skin cancer classification approach based on deep convolutional neural network, Cluster Computing (2024) 1--21.

\bibitem{madhu2022dscn}
G.~Madhu, A.~Govardhan, V.~Ravi, S.~Kautish, B.~S. Srinivas, T.~Chaudhary, M.~Kumar, Dscn-net: a deep siamese capsule neural network model for automatic diagnosis of malaria parasites detection, Multimedia Tools and Applications 81~(23) (2022) 34105--34127.

\bibitem{kumar2024application}
S.~A. Kumar, M.~K. Muchahari, S.~Poonkuntran, L.~S. Kumar, R.~K. Dhanaraj, P.~Karthikeyan, Application of hybrid capsule network model for malaria parasite detection on microscopic blood smear images, Multimedia Tools and Applications (2024) 1--27.

\bibitem{asif2024malaria}
H.~M. Asif, S.~H. Khan, T.~J. Alahmadi, T.~Alsahfi, A.~Mahmoud, Malaria parasitic detection using a new deep boosted and ensemble learning framework, Complex \& Intelligent Systems (2024) 1--17.

\bibitem{1111}
Z.~Liang, A.~Powell, I.~Ersoy, M.~Poostchi, K.~Silamut, K.~Palaniappan, P.~Guo, M.~A. Hossain, A.~Sameer, R.~J. Maude, J.~X. Huang, S.~Jaeger, G.~Thoma, Cnn-based image analysis for malaria diagnosis, in: 2016 IEEE International Conference on Bioinformatics and Biomedicine (BIBM), 2016, pp. 493--496.
\newblock \href {https://doi.org/10.1109/BIBM.2016.7822567} {\path{doi:10.1109/BIBM.2016.7822567}}.

\bibitem{rajaraman}
S.~Rajaraman, S.~K. Antani, M.~Poostchi, K.~Silamut, M.~A. Hossain, R.~J. Maude, S.~Jaeger, G.~R. Thoma, Pre-trained convolutional neural networks as feature extractors toward improved malaria parasite detection in thin blood smear images, PeerJ 6 (2018) e4568.

\bibitem{elangovan2021novel}
P.~Elangovan, M.~K. Nath, A novel shallow convnet-18 for malaria parasite detection in thin blood smear images: Cnn based malaria parasite detection, SN computer science 2~(5) (2021) 380.

\bibitem{marques2022ensemble}
G.~Marques, A.~Ferreras, I.~de~la Torre-Diez, An ensemble-based approach for automated medical diagnosis of malaria using efficientnet, Multimedia tools and applications 81~(19) (2022) 28061--28078.

\bibitem{quan2020effective}
Q.~Quan, J.~Wang, L.~Liu, An effective convolutional neural network for classifying red blood cells in malaria diseases, Interdisciplinary Sciences: Computational Life Sciences 12 (2020) 217--225.

\bibitem{murmu2024dlrfnet}
A.~Murmu, P.~Kumar, Dlrfnet: deep learning with random forest network for classification and detection of malaria parasite in blood smear, Multimedia Tools and Applications (2024) 1--23.

\bibitem{Vezakis2024-yg}
I.~A. Vezakis, K.~Georgas, D.~Fotiadis, G.~K. Matsopoulos, {EffiSegNet}: Gastrointestinal polyp segmentation through a pre-trained {EfficientNet-based} network with a simplified decoder (2024).
\newblock \href {http://arxiv.org/abs/2407.16298} {\path{arXiv:2407.16298}}.

\bibitem{Dumitru2023-gc}
R.-G. Dumitru, D.~Peteleaza, C.~Craciun, Using {DUCK-Net} for polyp image segmentation, Sci. Rep. 13~(1) (2023) 9803.

\bibitem{Zhou2023-tz}
L.~Zhou, Spatially exclusive pasting: A general data augmentation for the polyp segmentation, in: 2023 International Joint Conference on Neural Networks ({IJCNN}), IEEE, 2023.

\bibitem{Kato2022-wz}
S.~Kato, K.~Hotta, Adaptive {t-vMF} dice loss for multi-class medical image segmentation (2022).
\newblock \href {http://arxiv.org/abs/2207.07842} {\path{arXiv:2207.07842}}.

\bibitem{biswas2023polypsamtextguidedsam}
R.~Biswas, \href{https://arxiv.org/abs/2308.06623}{Polyp-sam++: Can a text guided sam perform better for polyp segmentation?} (2023).
\newblock \href {http://arxiv.org/abs/2308.06623} {\path{arXiv:2308.06623}}.
\newline\urlprefix\url{https://arxiv.org/abs/2308.06623}

\bibitem{lou2023caranet}
A.~Lou, S.~Guan, M.~Loew, Caranet: context axial reverse attention network for segmentation of small medical objects, Journal of Medical Imaging 10~(1) (2023) 014005--014005.

\bibitem{10.1007/978-3-030-59725-2_26}
D.-P. Fan, G.-P. Ji, T.~Zhou, G.~Chen, H.~Fu, J.~Shen, L.~Shao, \href{https://doi.org/10.1007/978-3-030-59725-2_26}{Pranet: Parallel reverse attention network for polyp segmentation}, in: Medical Image Computing and Computer Assisted Intervention – MICCAI 2020: 23rd International Conference, Lima, Peru, October 4–8, 2020, Proceedings, Part VI, Springer-Verlag, Berlin, Heidelberg, 2020, p. 263–273.
\newblock \href {https://doi.org/10.1007/978-3-030-59725-2_26} {\path{doi:10.1007/978-3-030-59725-2_26}}.
\newline\urlprefix\url{https://doi.org/10.1007/978-3-030-59725-2_26}

\bibitem{trinh2024samegsegmentmodelegde}
Q.-H. Trinh, H.-D. Nguyen, B.-T.~N. Ngoc, D.~Jha, U.~Bagci, M.-T. Tran, \href{https://arxiv.org/abs/2406.14819}{Sam-eg: Segment anything model with egde guidance framework for efficient polyp segmentation} (2024).
\newblock \href {http://arxiv.org/abs/2406.14819} {\path{arXiv:2406.14819}}.
\newline\urlprefix\url{https://arxiv.org/abs/2406.14819}

\bibitem{Fitzgerald2024-rk}
K.~Fitzgerald, J.~Bernal, A.~Histace, B.~J. Matuszewski, Polyp segmentation with the {FCB-SwinV2} transformer, IEEE Access (2024) 1--1.

\bibitem{Srivastava2022-do}
A.~Srivastava, D.~Jha, S.~Chanda, U.~Pal, H.~Johansen, D.~Johansen, M.~Riegler, S.~Ali, P.~Halvorsen, {MSRF-Net}: A multi-scale residual fusion network for biomedical image segmentation, IEEE J. Biomed. Health Inform. 26~(5) (2022) 2252--2263.

\bibitem{Zhou2018-cm}
Z.~Zhou, M.~M.~R. Siddiquee, N.~Tajbakhsh, J.~Liang, Unet++: A nested u-net architecture for medical image segmentation. In: Deep learning in medical image analysis and multimodal learning for clinical decision support, Springer, 2018.

\bibitem{Zhang2018-go}
Z.~Zhang, Q.~Liu, Y.~Wang, Road extraction by deep residual u-net, IEEE Geosci. Remote Sens. Lett. 15~(5) (2018) 749--753.

\bibitem{Valanarasu2021-ig}
J.~M.~J. Valanarasu, P.~Oza, I.~Hacihaliloglu, V.~M. Patel, Medical transformer: Gated axial-attention for medical image segmentation, in: Medical Image Computing and Computer Assisted Intervention -- {MICCAI} 2021, Springer International Publishing, Cham, 2021, pp. 36--46.

\bibitem{Chen2021-tp}
J.~Chen, Y.~Lu, Q.~Yu, X.~Luo, E.~Adeli, Y.~Wang, L.~Lu, A.~L. Yuille, Y.~Zhou, {TransUNet}: Transformers make strong encoders for medical image segmentation (2021).
\newblock \href {http://arxiv.org/abs/2102.04306} {\path{arXiv:2102.04306}}.

\bibitem{Valanarasu2022-nz}
J.~M.~J. Valanarasu, V.~M. Patel, {UNeXt}: {MLP-based} rapid medical image segmentation network, in: Lecture Notes in Computer Science, Springer Nature Switzerland, Cham, 2022, pp. 23--33.

\bibitem{10016712}
M.~Xu, K.~Huang, X.~Qi, A regional-attentive multi-task learning framework for breast ultrasound image segmentation and classification, IEEE Access 11 (2023) 5377--5392.
\newblock \href {https://doi.org/10.1109/ACCESS.2023.3236693} {\path{doi:10.1109/ACCESS.2023.3236693}}.

\bibitem{jin2023novel}
S.~Jin, S.~Yu, J.~Peng, H.~Wang, Y.~Zhao, A novel medical image segmentation approach by using multi-branch segmentation network based on local and global information synchronous learning, Scientific Reports 13~(1) (2023) 6762.

\bibitem{JIAO2024103202}
J.~Jiao, J.~Zhou, X.~Li, M.~Xia, Y.~Huang, L.~Huang, N.~Wang, X.~Zhang, S.~Zhou, Y.~Wang, Y.~Guo, \href{https://www.sciencedirect.com/science/article/pii/S1361841524001270}{Usfm: A universal ultrasound foundation model generalized to tasks and organs towards label efficient image analysis}, Medical Image Analysis 96 (2024) 103202.
\newblock \href {https://doi.org/https://doi.org/10.1016/j.media.2024.103202} {\path{doi:https://doi.org/10.1016/j.media.2024.103202}}.
\newline\urlprefix\url{https://www.sciencedirect.com/science/article/pii/S1361841524001270}

\bibitem{doiMalariaScreener}
H.~Yu, F.~Yang, S.~e.~a. Rajaraman, {M}alaria {S}creener: a smartphone application for automated malaria screening - {B}{M}{C} {I}nfectious {D}iseases (2020).
\newblock \href {https://doi.org/https://doi.org/10.1186/s12879-020-05453-1} {\path{doi:https://doi.org/10.1186/s12879-020-05453-1}}.

\bibitem{jha2020kvasir}
D.~Jha, P.~H. Smedsrud, M.~A. Riegler, P.~Halvorsen, T.~de~Lange, D.~Johansen, H.~D. Johansen, Kvasir-seg: A segmented polyp dataset, in: International Conference on Multimedia Modeling, Springer, 2020, pp. 451--462.

\bibitem{ALDHABYANI2020104863}
W.~Al-Dhabyani, M.~Gomaa, H.~Khaled, A.~Fahmy, \href{https://www.sciencedirect.com/science/article/pii/S2352340919312181}{Dataset of breast ultrasound images}, Data in Brief 28 (2020) 104863.
\newblock \href {https://doi.org/https://doi.org/10.1016/j.dib.2019.104863} {\path{doi:https://doi.org/10.1016/j.dib.2019.104863}}.
\newline\urlprefix\url{https://www.sciencedirect.com/science/article/pii/S2352340919312181}

\bibitem{Selvaraju_2019}
R.~R. Selvaraju, M.~Cogswell, A.~Das, R.~Vedantam, D.~Parikh, D.~Batra, \href{https://doi.org/10.1007%2Fs11263-019-01228-7}{Grad-{CAM}: Visual explanations from deep networks via gradient-based localization}, International Journal of Computer Vision 128~(2) (2019) 336--359.
\newblock \href {https://doi.org/10.1007/s11263-019-01228-7} {\path{doi:10.1007/s11263-019-01228-7}}.
\newline\urlprefix\url{https://doi.org/10.1007%2Fs11263-019-01228-7}

\bibitem{greenai}
R.~Schwartz, J.~Dodge, N.~A. Smith, O.~Etzioni, \href{https://doi.org/10.1145/3381831}{Green ai}, Commun. ACM 63~(12) (2020) 54–63.
\newblock \href {https://doi.org/10.1145/3381831} {\path{doi:10.1145/3381831}}.
\newline\urlprefix\url{https://doi.org/10.1145/3381831}

\end{thebibliography}


\end{document}